\documentclass[aps,prc,twocolumn,showpacs]{revtex4}
\usepackage{epsfig}
\newcommand{\ber}{\begin{eqnarray}}
\newcommand{\eer}{\end{eqnarray}}
\newcommand{\np}{{\rm N}_{\rm part}}
\newcommand{\pt}{p_{\rm T}}
\newcommand{\Et}{E_{\rm T}}

\begin{document}
\title{Deuteron-Nucleus Collisions in a Multi-Phase Transport Model}
\author{Zi-wei Lin}
\affiliation{Physics Department, The Ohio State University, 
Columbus, Ohio 43210}
\author{Che Ming Ko}
\affiliation{Cyclotron Institute and Physics Department, Texas A\&M University,
College Station, Texas 77843}
\date{\today}

\begin{abstract}
Using a multi-phase transport (AMPT) model, we study pseudo-rapidity 
distributions and transverse momentum spectra in deuteron-gold collisions 
at RHIC.  We find that final-state partonic and hadronic interactions 
affect the transverse momentum spectrum of protons more than those of
kaons or pions. Relative to p+p collisions at same center-of-mass
energy per nucleon pair, the effect of final-state interactions on the 
charged particle transverse momentum spectra in d+Au collisions is
much smaller than observed in experimental data, indicating that 
initial-state effects such as the Cronin effect are important. 
\end{abstract}
\maketitle

\section{introduction}

To study heavy ion collisions at the Relativistic Heavy Ion Collider 
(RHIC) at Brookhaven National Laboratory, in which a deconfined plasma
of quarks and gluons is expected to be formed, we have developed 
a multi-phase transport (AMPT) model that includes both final-state
partonic and hadronic interactions \cite{Zhang:ampt,Lin:2001cx,Lin:2001yd}. 
The model has been very useful for understanding various observables 
in Au+Au collisions at RHIC such as the rapidity and transverse momentum 
distributions of various particles \cite{Zhang:ampt,Lin:2001cx,Lin:2001yd} 
as well as charmonium \cite{Zhang:jpsi} and strangeness \cite{Pal:strange} 
productions. In particular, it allows one to study both thermal and chemical 
equilibration in the partonic and hadronic matter formed in these collisions. 
The AMPT model has also been extended to include the string melting mechanism 
\cite{Lin:2001zk,Lin:2002gc}, in which soft strings produced from 
initial nucleon-nucleon interactions are converted directly to partons, 
in order to explain the measured elliptic flow \cite{Ackermann:2001tr} 
and two-pion correlation functions \cite{Adler:2001zd} at RHIC.

In spite of its success, the AMPT model has many uncertainties
in its input physical parameters, particularly those related to 
the initial conditions introduced to the model. For example, the number
of initial minijet partons in the AMPT model, that is given by 
the hard processes from the HIJING model 
\cite{Wang:1991ht,Wang:1991us,Gyulassy:ew}, depends on the nuclear shadowing, 
i.e., the modification of the parton distributions in a nucleon when 
it is in a nucleus. As a result, the final particle multiplicity produced 
in relativistic heavy ion collisions is affected by nuclear shadowing 
\cite{Lin:2001cx,Lin:2001yd}. The production of high $\pt$ particles or 
hadrons made of heavy quarks, which is described by perturbative QCD 
processes, is even more sensitive to the nuclear shadowing effect. 
Furthermore, other nuclear effects such as the Cronin effect 
\cite{Cronin:zm}, which set in already in p-A collisions, need to be 
included. The uncertainties in the initial condition also exist in other 
transport models with partonic degrees of freedom 
\cite{Zhang:1997ej,Molnar:2000jh} as well as theoretical models such as the 
hydrodynamical model 
\cite{Kolb:2000fh,Huovinen:2001cy,Kolb:2001qz,Teaney:2001av} and
the QCD saturation model
\cite{Kharzeev:2000ph,Kharzeev:2001gp,Kharzeev:2002ei,EKRT,Eskola:2000xq}.

Since final-state interactions (FSI) are expected to be less important
in deuteron-gold collisions than in collisions between heavy nuclei, 
both the Cronin effect and the nuclear shadowing effect 
(e.g., through dilepton measurements \cite{Lin:1995pk}) can be 
better studied in these collisions. Improved knowledge on these
effects are useful for making reliable theoretical interpretations of 
the observations in heavy ion collisions at RHIC. In this paper, we
use the AMPT model to study the global observables in deuteron-gold 
collisions such as the pseudo-rapidity distributions and the transverse 
momentum spectra. Because of the small interaction volume in d+Au collisions, 
we use the default AMPT model, i.e., without the string melting mechanism, 
in the present study, as the initial energy density produced in these
collisions is expected to be small.

This paper is organized as follows. In Sec.~\ref{ampt}, we briefly
review the AMPT model. Results from the AMPT model on 
deuteron-gold collisions at RHIC are then shown in Sec.~\ref{results},     
with the charged particle pseudo-rapidity distributions and their 
centrality dependence given in Sec.~\ref{dndeta}, the transverse momentum 
spectra of different particles in Sec.~\ref{ptspectra}, 
and the effects due to modifications of the string fragmentation parameters 
and nuclear shadowing in Sec.~\ref{string}. A summary is then given in 
Sec.~\ref{summary}. Finally,  the effect of centrality selection on the 
centrality dependence of charged particle pseudo-rapidity distributions is 
discussed in the Appendix.

\section{The AMPT model}
\label{ampt}

The AMPT model is a hybrid model that consists of four components: 
the initial conditions, the parton cascade, the conversion from partonic 
to hadronic matter, and the hadron cascade. In the default AMPT model 
\cite{Zhang:ampt,Lin:2001cx,Lin:2001yd,Zhang:jpsi,Pal:strange}, 
the initial conditions are generated from the HIJING model 
\cite{Wang:1991ht,Wang:1991us,Gyulassy:ew} (version 1.383 for this study), 
which usually uses a Woods-Saxon radial shape for the colliding nuclei and 
introduces a parameterized nuclear shadowing function that depends on the 
impact parameter of the collision \cite{Wang:1991ht}. Interactions among 
minijet partons, which are produced from initial hard nucleon-nucleon 
interactions, are modeled by the Zhang's parton cascade (ZPC) 
\cite{Zhang:1997ej}. After partons stop interacting, they recombine with 
their parent strings, which are produced from initial soft nucleon-nucleon 
interactions, and fragment to hadrons according to the Lund string 
fragmentation model \cite{Sjostrand:1993yb,Andersson2}. Dynamics of  
resulting hadronic matter is then described by a hadronic cascade based 
on the relativistic transport (ART) model \cite{Li:1995pr}. Final hadronic 
observables including contributions from the strong decay of resonances 
are determined when the hadronic matter freezes out.
 
In this study, we use the Hulthen wave function \cite{hulthen} to model
the structure of a deuteron \cite{hj1382}:
\ber
u (r)= C e^{-\alpha r} \left ( 1-e^{-\mu r} \right ),
\eer
where $r$ represents the relative distance between the proton and neutron 
in the deuteron, the normalization constant $C$ is determined from 
$\int_0^\infty u(r)^2 dr=1$, and the small $D$-wave contribution ($\sim$6\%) 
to the deuteron wave function is neglected.  With 
$\alpha=(4.38~{\rm fm})^{-1}$ and $\mu=(1.05~{\rm fm})^{-1}$, 
the root-mean-square radius of the deuteron, $\sqrt {\langle r^2 \rangle}/2$, 
is 2.0 fm, consistent with the measured value. 
 
\section{AMPT results on deuteron-gold collisions}
\label{results}

We have studied deuteron-gold collisions at $\sqrt {s_{NN}}=200$ GeV with 
both deuteron and gold having the same energy of $100$ GeV per nucleon. 
In this study, ``minimum bias'' d+Au events are defined as those within 
the impact parameter range between $0$ and $12$ fm, and they are separated 
into different centrality bins.

\subsection{Pseudo-rapidity distributions of charged particles}
\label{dndeta}

\begin{figure}[htb]
\centerline{\epsfig{file=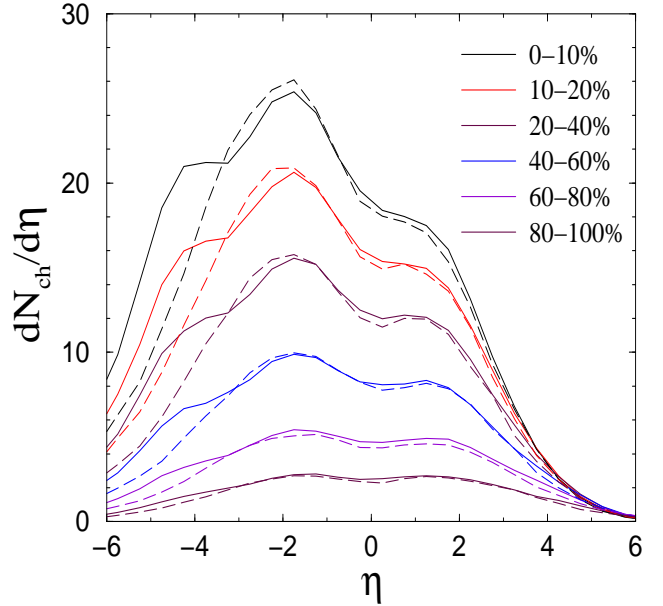,width=3.3in,height=3.3in}}
\caption{Pseudo-rapidity distributions of charged particles in d+Au 
collisions at $\sqrt {s_{NN}}=200$ GeV with centralities determined from 
\protect{$\np$}. Solid and dashed curves represent results from the AMPT
model and the HIJING model (without quenching), respectively.}
\label{dnchde}
\end{figure}

Fig.~\ref{dnchde} shows the pseudo-rapidity distributions of charged 
particles in the six centrality bins of 0-10\%, 10-20\%, 20-40\%, 40-60\%, 
60-80\%, and 80-100\%.  Here, the centrality is determined according to the 
value of $\np$, i.e., the total number of participants in both deuteron 
and gold nuclei, in each event. The pseudo-rapidity in this study is 
evaluated in the nucleon-nucleon center-of-mass frame, and negative 
rapidities correspond to the fragmentation region of the gold nucleus. 
Solid curves are results from the AMPT model, while dashed curves are 
those from the HIJING model without quenching. Note that the AMPT model 
without final-state partonic and hadronic interactions is equivalent to 
the HIJING model without jet quenching if the popcorn mechanism, controlled 
mainly by parameters MSTJ(12) and PARJ(5) in the PYTHIA/JETSET program 
\cite{Sjostrand:1993yb}, is treated in the same way. It is seen that the 
asymmetry in the pseudo-rapidity distributions, e.g., by comparing values of 
$dN_{ch}/d\eta$ at $\eta=-2$ and $\eta=+2$, decreases as collisions 
become less central, and the pseudo-rapidity distribution for the 
most peripheral bin is almost symmetric. As expected, final-state 
interactions in d+Au collisions have a smaller effect on charged particle 
pseudo-rapidity distributions than in the case of central heavy ion 
collisions at RHIC \cite{Lin:2001cx,Lin:2001yd}.

We note that the appearance of small bumps around $\eta\sim -4$ 
is mainly due to interactions of produced particles with initial incoming 
nucleons in the gold nucleus, which has an initial half width of 
$R_{Au}/\gamma$ in the longitudinal direction with $R_{Au}$ being 
the hard-sphere radius of the gold nucleus and $\gamma$ denoting the Lorentz 
boost factor of a nucleon in the nucleon-nucleon center-of-mass frame. 
If the initial half width is reduced by a factor of 5, the charged particle 
pseudo-rapidity distributions would be smooth around $\eta\sim -4$
and also have slightly lower values in the region of $-6<\eta<-3$.

\begin{figure}[htb]
\centerline{\epsfig{file=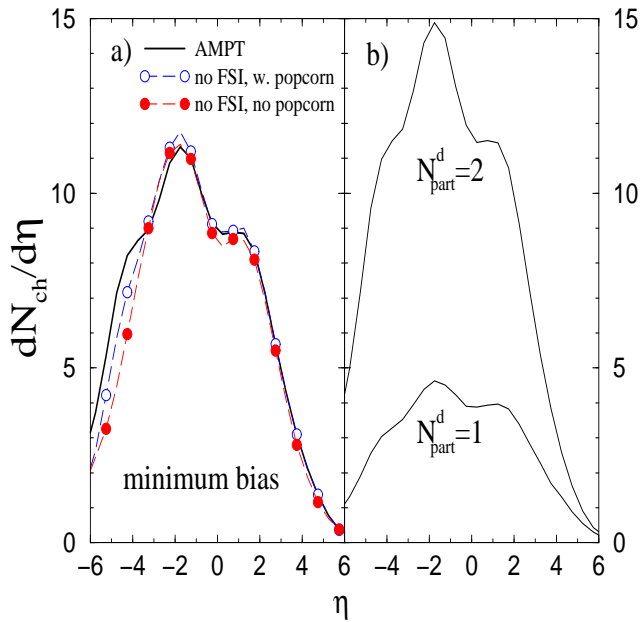,width=3.3in,height=3.3in}}
\caption{Pseudo-rapidity distributions of charged particles 
a) for ``minimum bias'' events when different interactions are included and 
b) for events from AMPT with $N^{\rm d}_{\rm part}=2$ 
and with $N^{\rm d}_{\rm part}=1$.}
\label{dnchde-mb}
\end{figure}

Fig.~\ref{dnchde-mb}a) shows the pseudo-rapidity distributions of charged 
particles for ``minimum bias'' d+Au events. Solid curves are results 
from the AMPT model, dashed curves with open circles are AMPT results 
without final-state interactions, and dashed curves with filled circles
are those without both final-state interactions and the popcorn mechanism 
for baryon-antibaryon production \cite{Zhang:ampt,Lin:2001cx,Lin:2001yd}. 
We see that both final-state interactions and the popcorn mechanism 
broaden the pseudo-rapidity distributions, especially in the fragmentation 
region of the gold nucleus, leading thus to a moderate increase of
the particle multiplicity in the region of $-6<\eta<-3$. We note that
without final-state interactions and the popcorn mechanism, results
from the AMPT model are equivalent to those from the HIJING model 
(without jet quenching) as same values of $a$ and $b$ in
Eq.(\ref{lund}) are used.  

One can separate events according to the number of participants from the 
deuteron, i.e., $N^{\rm d}_{\rm part}$. As an example, 
Fig.~\ref{dnchde-mb}b) shows the pseudo-rapidity distributions of charged 
particles from the AMPT model for events with one or two nucleon(s) 
from the deuteron that are involved in primary collisions. About 2/3 of
``minimum bias'' events from the AMPT model have $N^{\rm d}_{\rm part}=2$ 
with average values of $N^{\rm Au}_{\rm part}$ at $10.4$ and the mean 
impact parameter at $4.7$ fm. On the other hand, for events with 
$N^{\rm d}_{\rm part}=1$ the average value of $N^{\rm Au}_{\rm part}$
is $2.8$ and the mean impact parameter is $7.8$ fm. 

\begin{figure}[htb]
\centerline{\epsfig{file=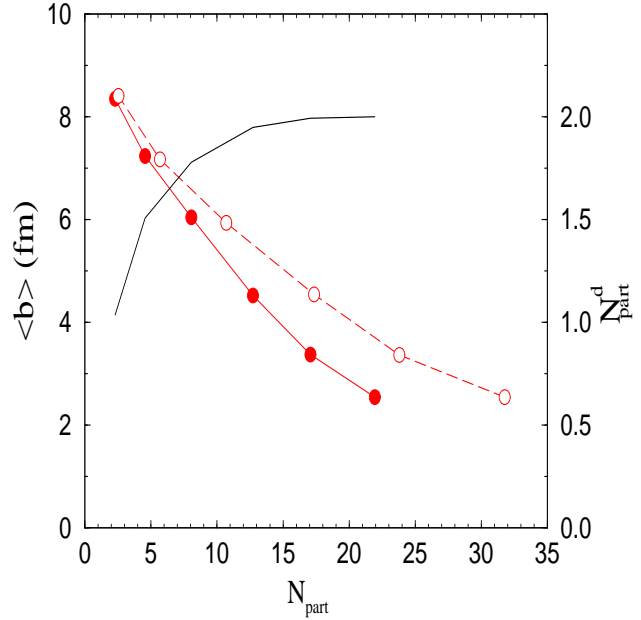,width=3.3in,height=3.3in}}
\caption{The average value of impact parameter (in fm) 
and $N^{\rm d}_{\rm part}$ (solid curve) 
as a function of the total $\np$. See text for details.}
\label{npart-pc}
\end{figure}

In Fig.~\ref{npart-pc}, we show by the solid curve with circles the 
average impact parameter as a function of the average value of $\np$ 
at each centrality bin. Here, the number of participants is defined as the 
number of initial (projectile and target) nucleons involved in {\it primary} 
collisions, i.e., not including those initial nucleons that interact 
with produced particles in the final state. The solid curve with no
symbols in Fig.~\ref{npart-pc} represents $N^{\rm d}_{\rm part}$.
We see that $N^{\rm d}_{\rm part}$ increases toward the value of 2
rather quickly from peripheral to central collisions. The dashed curve 
with open circles shows the number of participating nucleons that also 
include initial nucleons that are involved in final-state interactions. 
It is seen that final-state interactions increase the number of 
participating nucleons by more than 40\% for the most central events 
(0-10\% centrality bin). However, since the energies involved in 
final-state interactions are much lower than those in the primary collisions, 
they contribute much less to particle production. In this study, we
thus only include initial nucleons that are involved in primary
collisions for determining $\np$.

\begin{figure}[htb]
\centerline{
\epsfig{file=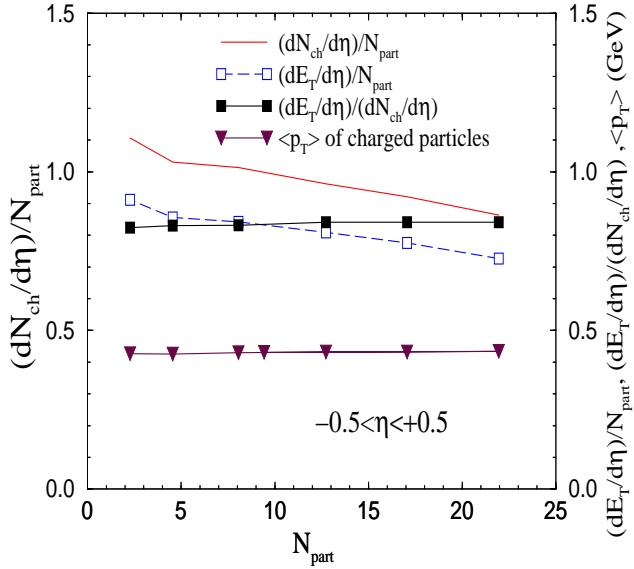,width=3.3in,height=3.3in}}
\caption{Centrality dependences of charged particle multiplicity and 
transverse energy (in GeV) per participant, the mean transverse energy 
(in GeV) per charged particle, and the mean $\pt$ (in GeV$/c$) of charged 
particles at mid-pseudo-rapidity.}
\label{npart-eta0}
\end{figure}

The centrality dependence of particle multiplicity and transverse energy at 
mid-pseudo-rapidities ($-0.5\!<\eta<\!0.5$) are shown Fig.~\ref{npart-eta0}. 
The solid curve and the dashed curve with open squares represent, 
respectively, $dN_{ch}/d\eta/\np$ and $d\Et/d\eta/\np$, where $\Et$ 
includes the contribution from neutral particles.  In this study, 
$\Et=E \sin \theta$ with $\theta$ being the polar angle,  
and $E$ is defined as the kinetic energy for baryons, the total energy 
including the mass for anti-baryons, and the total energy for all
other particles. We observe that both the particle multiplicity and transverse 
energy per participant gradually decrease with increasing $\np$. However, we
shall see in Sec.~\ref{string} and the Appendix that this moderate decrease 
with centrality will change when centralities are determined differently
or when the string fragmentation function is modified in more central 
d+Au collisions. Fig.~\ref{npart-eta0} also shows that the ratio of
the transverse energy and particle multiplicity per participant (solid
curve with filled squares), i.e., the average transverse energy per 
charged particle, is rather flat. Also shown by the solid curve with
triangles is the mean transverse momentum of charged particles, which 
is seen to change little with $\np$ as well.

\begin{figure}[htb]
\centerline{\epsfig{file=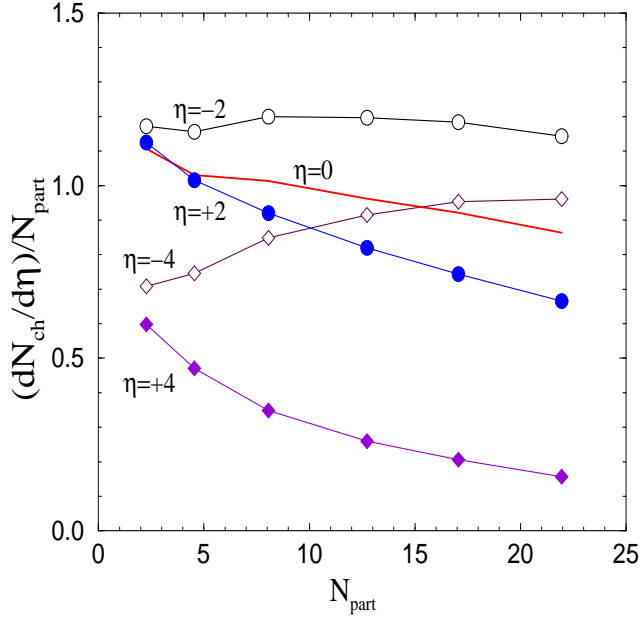,width=3.3in,height=3.3in}}
\caption{Centrality dependence of charged particle multiplicity per 
participant at different pseudo-rapidities.}
\label{nch-5eta}
\end{figure}

The centrality dependence of the charged particle multiplicity per 
participant at different pseudo-rapidities is shown in Fig.~\ref{nch-5eta}, 
where the value at each $\eta$ represents the average value 
within the pseudo-rapidity range of $\eta \pm 0.5$. We see an 
increase of this quantity at the backward pseudo-rapidity $\eta=-4$  
(curve with open diamonds) and a fast decrease at the forward 
pseudo-rapidity $\eta=+4$ (curve with filled diamonds). Furthermore,
the decrease with centrality becomes stronger as the pseudo-rapidity
changes from negative values to more positive values. These are consistent 
with the picture that multiple interactions in the gold nucleus push 
particle productions toward the negative rapidity region.  Since we find 
that the mean transverse energy per charged particle at a given 
pseudo-rapidity does not change much with centrality, the centrality 
dependence of $(d\Et/d\eta)/\np$, i.e., the transverse energy per participant, 
has a similar behavior at each pseudo-rapidity as $(dN_{ch}/d\eta)/\np$. 
  
\subsection{Transverse momentum spectra}
\label{ptspectra}

\begin{figure}[htb]
\centerline{\epsfig{file=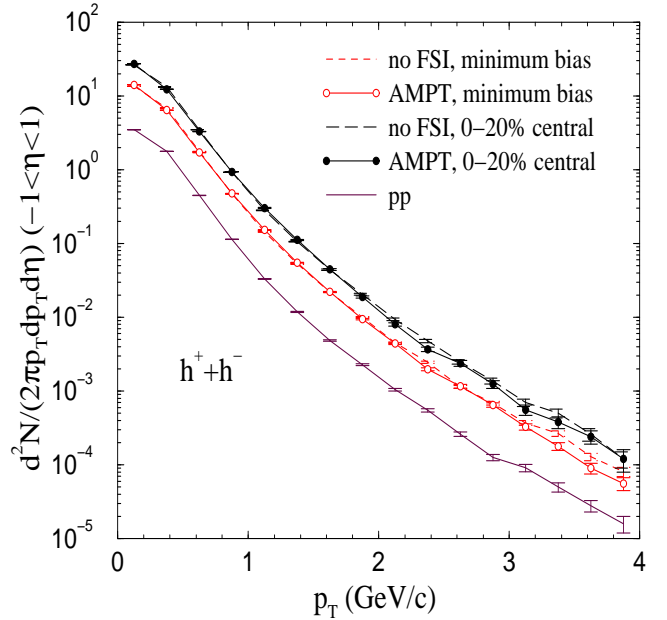,width=3.3in,height=3.3in}}
\caption{Transverse momentum spectra of charged particles within $-1<\eta<1$ 
for d+Au collisions from AMPT with (solid curves with circles) or without 
(dashed) final-state interactions. The solid curve is the AMPT result 
for p+p collisions at the same center-of-mass energy per nucleon pair.}
\label{chpt}
\end{figure}

The transverse momentum spectra of charged particles in the pseudo-rapidity 
range of $-1<\eta<1$ for both ``minimum bias'' and  0-20\% central d+Au 
collisions are shown in Fig.~\ref{chpt}. Solid curves with circles 
are results from the AMPT model while dashed curves are those without
final-state partonic and hadronic interactions. For comparisons, the 
AMPT results for minimum bias p+p collisions are shown by the solid curve. 
In all transverse momentum distributions and their ratios shown later,
we include the statistical errors in the AMPT model, which become
large at moderately high $\pt$.

\begin{figure}[htb]
\centerline{\epsfig{file=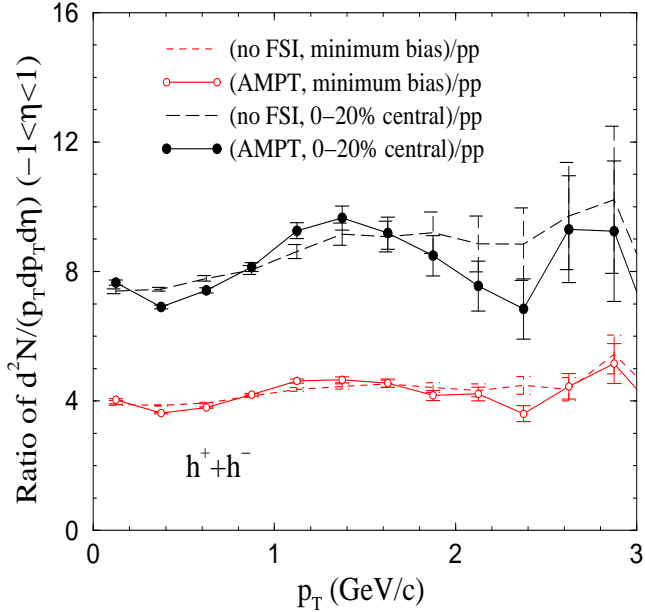,width=3.3in,height=3.3in}}
\caption{Ratios of the charged particle transverse momentum spectra 
for d+Au collisions over the spectra for p+p collisions 
from AMPT with (solid) or without (dashed) final-state interactions.}
\label{chpt-ratio}
\end{figure}

To better examine the distributions shown in Fig.~\ref{chpt},
ratios of the $\pt$ spectra of charged particles from d+Au collisions over 
those from minimum bias p+p collisions are shown in Fig.~\ref{chpt-ratio}.
We see that, even without final-state partonic and hadronic interactions, 
the ratios (dashed curves) are not flat but instead tend to increase 
with $\pt$ in the range of $0-3$ GeV$/c$ shown in the figure. This behavior
is mainly a result of different scalings for hard and soft processes 
with respect to the number of binary collisions in the initial
conditions from the HIJING model. While particle production at high 
enough $\pt$ scales with the number of initial binary collisions, 
low momentum particles have a weaker dependence.
Final-state interactions further modify the ratios of $\pt$ spectra, 
as seen from comparisons between the solid and dashed curves.

Recently, the transverse momentum spectra of charged particles 
in d+Au collisions have been measured at RHIC \cite{rhic-dAu}. 
Compared with the experimental data at $0<\pt<3$ GeV$/c$, 
the $\pt$ dependence of the ratios from the AMPT calculations shown in 
Fig.~\ref{chpt-ratio}, either with or without final-state interactions, 
are much weaker. Thus, interactions in the final state are not the main reason 
for the observed strong $\pt$ dependence of the ratios of transverse
momentum spectra, suggesting that initial-state effects such as parton 
momentum broadening due to the Cronin effect \cite{Cronin:zm,Vitev:2003xu} 
(not included in the AMPT model so far) are important in d+Au collisions. 

\begin{figure}[htb]
\centerline{\epsfig{file=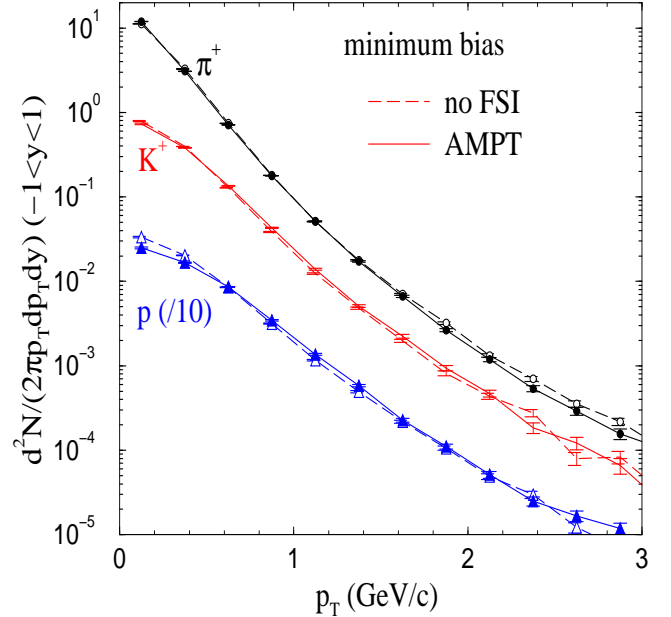,width=3.3in,height=3.3in}}
\caption{Transverse momentum spectra of $\pi^+,K^+$ and protons 
from AMPT with (solid) or without (dashed) final-state 
interactions for ``minimum-bias'' d+Au collisions. 
The proton spectra are scaled down by a factor of 10.}
\label{idpt-mb}
\end{figure}

For identified particles such as $\pi^+, K^+$ and protons,
their transverse momentum spectra within the rapidity range of $-1<y<1$
from ``minimum bias'' d+Au collisions are shown in Fig.~\ref{idpt-mb}, 
with the proton spectra scaled down by a factor of 10.  Solid and
dashed curves represent AMPT results with and without final-state 
interactions, respectively.  Compared with results from the AMPT model
on central heavy ion collisions, e.g., Fig.~2 of Ref.~\cite{Lin:2001yd} 
for central Pb+Pb collisions at SPS, the effect of final-state interactions 
on the transverse momentum spectra in d+Au collisions are much weaker.

\begin{figure}[htb]
\centerline{\epsfig{file=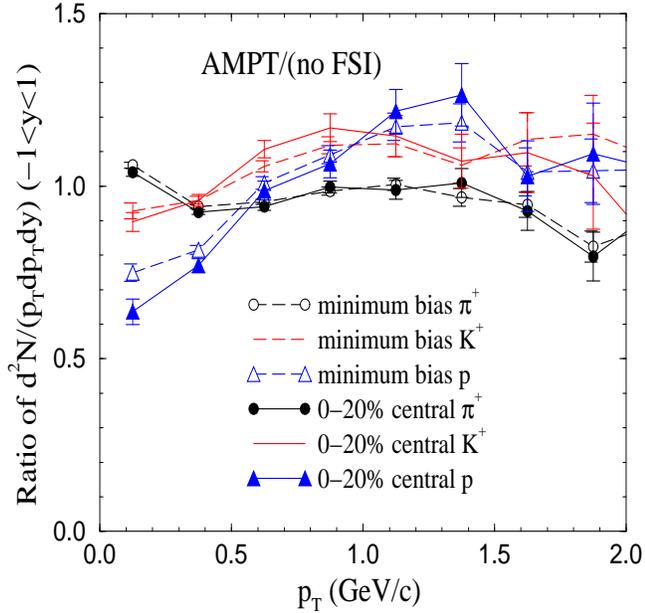,width=3.3in,height=3.3in}}
\caption{Ratios of the transverse momentum spectra of $\pi^+$, $K^+$, 
and protons from AMPT over those from the AMPT model without
final-state interactions for ``minimum bias'' (dashed) and 0-20\%
central (solid) d+Au collisions.}
\label{idpt-ratio-nofsi}
\end{figure}

Fig.~\ref{idpt-ratio-nofsi} shows the ratios of the $\pt$ spectra from AMPT 
over those from the AMPT model without final-state interactions. 
Dashed curves represents the ratios of the ``minimum bias'' spectra shown in 
Fig.~\ref{idpt-mb}, while solid curves represent the ratios of the 
spectra from 0-20\% d+Au collisions. We see that final-state interactions 
modify the $\pt$ spectra of $\pi, K$ and protons differently. Proton 
spectra show the largest FSI effect with low $\pt$ protons shifting to 
higher $\pt$, and the ratio for 0-20\% central events has a stronger 
$\pt$ dependence than that for ``minimum bias'' events. Both features 
are consistent with the existence of some transverse flow in d+Au collisions. 
We find that the average number of partonic collisions per parton 
is about 0.06 (with $3$ mb parton scattering cross section) 
and the average number of hadronic collisions per produced hadron is 
about 1 in ``minimum bias'' d+Au collisions from the AMPT model, compared 
with about 1 partonic collision per parton and 8 hadronic collisions
per hadron in central Au+Au collisions at $200A$ GeV. As a result, the 
transverse flow due to final-state interactions in d+Au collisions is
much weaker than in central heavy ion collisions. We also note that in 
the default AMPT model used in this study, final-state interactions 
among partons are less important than those among hadrons as only
minijet partons are included in the parton cascade.

\begin{figure}[htb]
\centerline{\epsfig{file=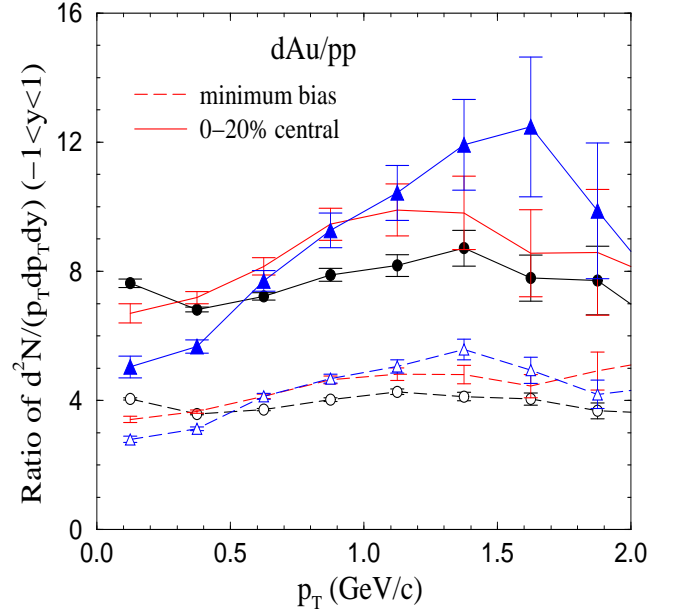,width=3.3in,height=3.3in}}
\caption{Ratios of the transverse momentum spectra from AMPT for d+Au 
collisions over those for p+p collisions. Solid and dashed curves 
correspond to 0-20\% central and ``minimum-bias'' d+Au collisions, 
respectively.}
\label{idpt-ratio-pp}
\end{figure}

Ratios of the $\pt$ spectra in d+Au collisions over those in p+p 
collisions from the AMPT model are shown in Fig.~\ref{idpt-ratio-pp}.
Similar to Fig.~\ref{idpt-ratio-nofsi}, we observe the strongest $\pt$ 
dependence for protons. Also, ratios of the $\pt$ spectra of kaons 
have a stronger $\pt$ dependence than those of pions. 
We note that, since the initial-state parton broadening 
due to the Cronin effect \cite{Cronin:zm} has not yet been included 
in the AMPT model, the $\pt$ dependence of these ratios may get even stronger 
if the Cronin effect is taken into account.

\subsection{Effects of the string fragmentation parameters and 
nuclear shadowing}
\label{string}

In the default AMPT model, hadron production after the partonic phase 
is described by the Lund string fragmentation model. 
In this model, the longitudinal momentum distribution
of a hadron with transverse mass $m_{\perp}$ produced from the
string fragmentation is given by the following symmetric splitting function 
\cite{Andersson2}: 
\ber
f(z) \propto z^{-1} (1-z)^a \exp (-b~m_{\perp}^2/z), 
\label{lund}
\eer
where $z$ denotes the light-cone momentum fraction of the produced 
hadron with respect to that of the fragmenting string. 
In the HIJING model, which reproduces the experimental 
charged particle multiplicities in high energy p+p and $p\bar p$ collisions,  
the default values of $a=0.5$ and $b=0.9$ GeV$^{-2}$ are used. 
However, in order to reproduce the rapidity distributions 
of charged particles in central Pb+Pb collisions at the CERN-SPS energy 
using the AMPT model, we find that values of these 
parameters need to change to $a=2.2$ and $b=0.5$ GeV$^{-2}$ 
\cite{Lin:2001cx,Lin:2001yd}. 
The increase of $a$ and decrease of $b$ relative to the default
values in the HIJING model soften the splitting function and thus 
enhance the total charged particle multiplicity by about 20\% in 
central Pb+Pb collisions at SPS. The change of these two parameters 
can perhaps be attributed to the modification of string fragmentation 
when multiple strings are produced and overlap in heavy ion collisions,  
although it is not known yet how these parameters should be modified
in different colliding systems. Since peripheral d+Au collisions are 
expected to behave similarly as p+p collisions, and even in central 
d+Au collisions the string density is much lower than in central heavy 
ion collisions, we have assumed that the parameters $a$ and $b$ of 
Eq.~(\ref{lund}) for deuteron-gold collisions have the same values as 
in p+p collisions, i.e., they take the default values in the HIJING model. 

\begin{figure}[htb]
\centerline{\epsfig{file=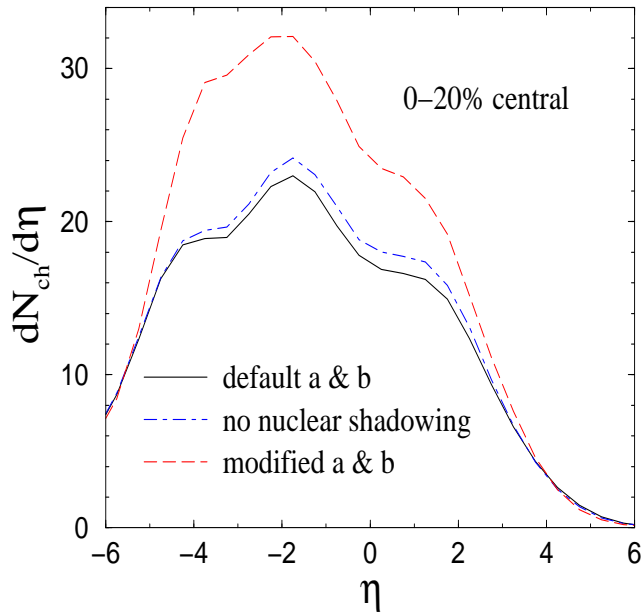,width=3.3in,height=3.3in}}
\caption{Charged particles pseudo-rapidity distributions for 
the most central 20\% deuteron-gold collisions from AMPT
with default values of $a$ and $b$ parameters (solid), with nuclear 
shadowing turned off (dot-dashed), or with modified $a$ and $b$ values 
as in heavy ion collisions (dashed).} 
\label{nch-ab-shad}
\end{figure}

However, since the values of $a$ and $b$ parameters need to be modified 
in order to describe the total multiplicity in central heavy ion collisions, 
it is still possible that these parameters 
are also modified in non-peripheral deuteron-gold collisions. 
If in central deuteron-gold collisions the parameters $a$ and $b$ 
already have the modified values as in heavy ion collisions, 
the pseudo-rapidity distribution for the 20\% most central deuteron-gold 
collisions (selected from $\np$) 
would be given by the dashed curve in Fig.~\ref{nch-ab-shad}, 
which shows a substantial increase compared with the previous AMPT result 
obtained with default $a$ and $b$ values in the HIJING model (solid curve). 
Comparison of these results for deuteron-gold collisions 
with upcoming experimental results at RHIC will thus allow us to 
learn how the string fragmentation process is modified in different 
colliding systems or centralities. 

\begin{figure}[htb]
\centerline{\epsfig{file=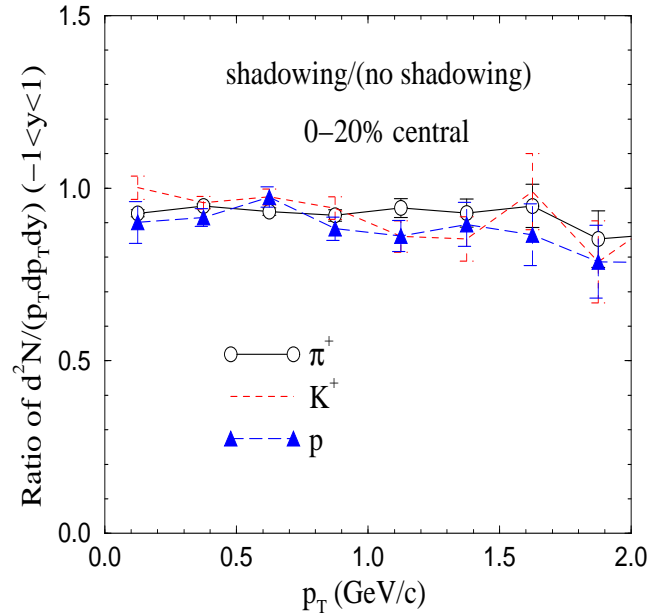,width=3.3in,height=3.3in}}
\caption{Ratios of the transverse momentum spectra of $\pi^+, K^+$ and protons
from AMPT with nuclear shadowing over those without nuclear shadowing 
for 0-20\% central d+Au events.}
\label{idpt-ratio-shad}
\end{figure}

When the nuclear shadowing effect is turned off in the AMPT model, the 
pseudo-rapidity distribution for the 20\% most central deuteron-gold collisions
is shown by the dot-dashed curve in Fig. \ref{nch-ab-shad}, which is not much 
different from the solid curve obtained with the nuclear shadowing.
Thus, pseudo-rapidity distributions of charged particles in deuteron-gold 
collisions are not very sensitive to the nuclear shadowing effect. 
This is different from that seen in collisions between heavy nuclei, where 
the effect is much larger as nuclear shadowing affects the production 
of minijet partons, which scales with the number of binary collisions. 
The effects of nuclear shadowing on $\pt$ spectra of 
$\pi^+, K^+$ and protons, shown in Fig.~\ref{idpt-ratio-shad} for 
0-20\% central d+Au collisions, are also relatively small and 
do not have a strong $\pt$ dependence within $0<\pt<3$ GeV$/c$.
Note that, although the nuclear shadowing effect is unimportant for global 
observables in small colliding systems such as deuteron-gold collisions, 
it affects significantly observables which are dominated by 
partonic interactions such as open charm production
\cite{Lin:1995pk} or the yield of high $\pt$ particles. 

\section{summary}
\label{summary}

Using a multi-phase transport (AMPT) model that includes both final-state 
partonic and hadronic interactions, we have studied the pseudo-rapidity 
distributions of charged particles, their centrality dependence, and 
transverse momentum spectra of different particles in deuteron-gold 
collisions at $\sqrt {s_{NN}}=200$ GeV. Due to the asymmetry of these 
collisions, the centrality dependence of charged particle multiplicity 
per participant is very different at different pseudo-rapidities, and 
it goes from increasing with centrality at the backward
pseudo-rapidity region (the fragmentation region of the gold nuclei) 
to decreasing with centrality at the forward region. 
The charged particle pseudo-rapidity distribution in central deuteron-gold 
collisions is also sensitive to the values of the parameters used 
in the string fragmentation function. Furthermore, we find that, 
although final-state partonic and hadronic interactions modify 
the transverse momentum spectra of charged particles in d+Au collisions 
relative to scaled p+p collisions, the $\pt$ dependence of the modification 
due to final-state interactions is much weaker than observed at RHIC. 
Thus, initial-state effects such as parton momentum broadening due to the 
Cronin effect are important in deuteron-gold collisions. However, we find that 
final-state interactions have a much stronger effect on the $\pt$ spectra
of protons than those of kaons or pions. Comparison of these
predictions with the experimental data will thus help us to learn more 
about initial-state effects on transverse momentum spectra and to
study whether string fragmentation is modified in deuteron-gold 
collisions as in central heavy ion collisions. 

\appendix*
\section{Effects of different centrality selections}

To extract useful information from the centrality dependence of 
global observables, it is helpful to study how the criteria of 
centrality selection affects the centrality dependence. The number of 
participants and the impact parameter are known in theoretical models, 
but they cannot be directly measured in experiments. 
Instead, experimental centrality selections usually involve  
cutting on the charged particle multiplicity or the transverse energy. 
Thus, in addition to separating events into different centrality bins
using the total number of participants, we have also tried three other 
methods, using the impact parameter, the number of charged particles 
within $-1\!<\eta<\!1$, and the total number of charged particles. 
For example, when the total number of charged particles 
is used for the centrality selection, we order all 35,000 events from 
the AMPT model by the total number of charged particles in each event.  
The centrality bin of, say 0-10\%, then consists of 
the first 3,500 events in that ordered list.

\begin{figure}[htb]
\centerline{\epsfig{file=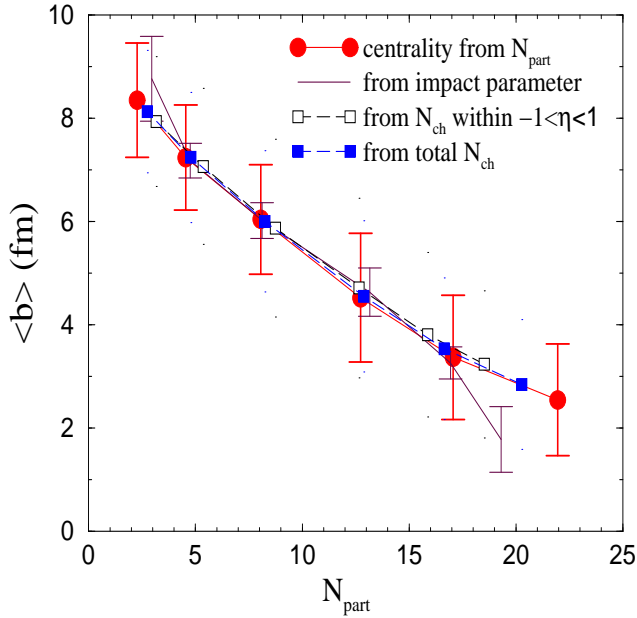,width=3.3in,height=3.3in}}
\caption{The average value of impact parameter as a function of 
$\np$ with different criteria for the centrality selection, 
together with the RMS width of the impact parameter in each centrality bin
for two of the curves.} 
\label{b-4way}
\end{figure}

In Fig.~\ref{b-4way}, we plot the average value of impact parameters 
versus the average value of $\np$ at six centrality bins of 
$0-10-20-40-60-80-100\%$, that are selected using these four different 
methods. The solid curve with circles corresponds to the 
centrality selection from $\np$, and it thus has the largest value of 
$\np$ for the most central bin of events and the smallest value of 
$\np$ for the most peripheral bin of events.  On the other hand, 
the centrality selection from the impact parameter, shown by the solid 
curve with no symbols, has the smallest value of $\langle{\rm b}\rangle$ 
for the most central bin of events and the largest value of 
$\langle {\rm b} \rangle$ for the most peripheral bin of events. 
When the number of charged particles is used to determine the centrality, 
the correlation between the centrality and $\np$ is weaker than 
in the case of using $\np$. This is especially true when only part of 
the phase space is included, as using $N_{ch}$ within $-1\!<\eta<\!1$ 
for the centrality selection (curve with open squares) leads to a
narrower range of $\np$ than using the total number of charged particles 
for the centrality selection (curve with filled squares).
The error bars shown for two of the curves in Fig.~\ref{b-4way}  
(the curves from using $\np$ or impact parameter for the centrality selection) 
correspond to the root-mean-square widths of the impact parameters in 
each centrality bin, and we see that the width in the case of 
using $\np$ for determining the centrality is be quite large (around 1.2 fm) 
for these centrality bins.  We also find that, when $N_{ch}$ within 
$-1\!<\eta<\!1$ or the total $N_{ch}$ is used for the centrality selection, 
these widths are even larger (between 1.3 and 1.8 fm), indicating that 
the correlation between $N_{ch}$ and the impact parameter 
is weaker than that between $\np$ and the impact parameter.

\begin{figure}[htb]
\centerline{\epsfig{file=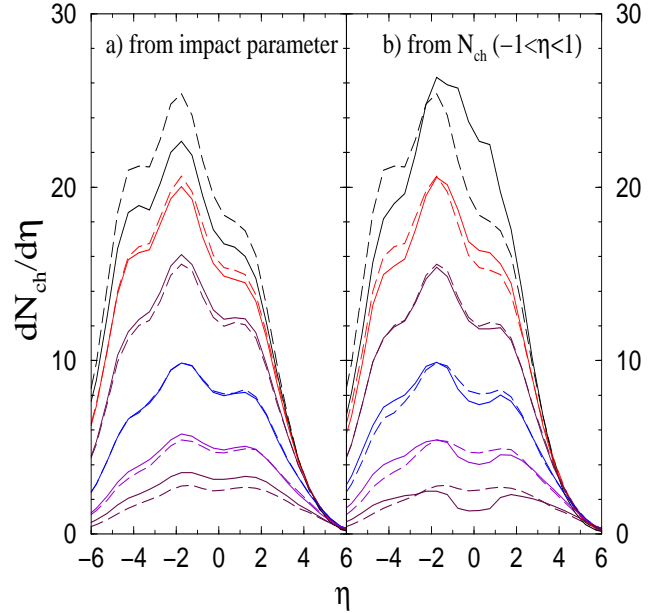,width=3.3in,height=3.3in}}
\caption{Charged particle pseudo-rapidity distributions at six centrality bins 
of $0-10-20-40-60-80-100\%$ when the centrality is determined from 
a) the impact parameter or b) $N_{ch}$ within $-1\!<\eta<\!1$. Dashed curves 
represent results when the centrality is determined from $\np$.}
\label{dnchde-w2}
\end{figure}

\begin{figure}[htb]
\centerline{\epsfig{file=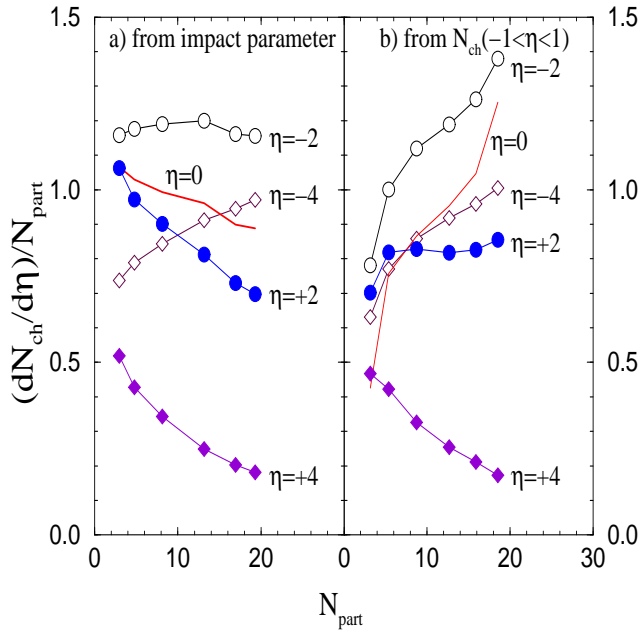,width=3.3in,height=3.3in}}
\caption{Centrality dependence of charged particle multiplicity with
the centrality determined from a) the impact parameter or b) $N_{ch}$ 
within $-1\!<\eta<\!1$.}
\label{npart-nch-w2}
\end{figure}

Different centrality selections also lead to differences in both the 
shape and magnitude of the pseudo-rapidity distributions at 
the same centrality bin. Solid curves in Fig.~\ref{dnchde-w2}a and 
\ref{dnchde-w2}b show the pseudo-rapidity distributions of charged 
particles using, respectively, the impact parameter and $N_{ch}$ 
within $-1\!<\eta<\!1$ for determining the centrality. The distributions 
corresponding to using $\np$ for the centrality selection are also 
shown by the dashed curves for comparison. From Fig.~\ref{dnchde-w2}a, 
we find that $dN_{ch}/d\eta$ at $\eta=0$ grows slower with centrality when 
the impact parameter instead of $\np$ is used for the centrality selection. 
However, it is interesting to see, from comparing Fig.~\ref{npart-nch-w2}a 
with Fig.~\ref{nch-5eta}, that the centrality dependences of 
$(dN_{ch}/d\eta)/\np$ for the two cases are quite similar. This is due 
to the compensating effect of slower growth of $\np$ with centrality 
when the impact parameter is used to select the centrality 
(see Fig.~\ref{b-4way}).  We note that in relativistic heavy ion 
collisions where the multiplicity is much higher, these different 
methods for the centrality selection lead to quite similar results, 
contrary to those shown in Figs.~\ref{b-4way} and ~\ref{dnchde-w2} 
for d+Au collisions.

Fig.~\ref{npart-nch-w2}b shows the centrality dependence of charged 
particle multiplicity per participant at different pseudo-rapidities 
when $N_{ch}$ within $-1\!<\eta<\!1$ is used for the centrality selection.  
Comparing with Fig.~\ref{nch-5eta}, we find that the centrality 
dependence at $\eta=-4$ or $\eta=+4$ is similar to the case where 
$\np$ determines the centrality, but the centrality dependence at 
$\eta=-2, 0$ or $+2$ is totally different. For example, while in 
Fig.~\ref{nch-5eta} (and in Fig.~\ref{npart-nch-w2}a) the curve 
of $(dN_{ch}/d\eta)/\np$ at $\eta=0$ shows a small decrease with 
centrality, the corresponding curve in Fig.~\ref{npart-nch-w2}b 
increases significantly with centrality. We note that these differences 
from Fig.~\ref{nch-5eta} exist even when the total 
number of $N_{ch}$ is used for selecting the centrality 
(instead of $N_{ch}$ within $-1\!<\eta<\!1$). Part of this large difference 
in the centrality dependence is due to the stronger increase of 
$dN_{ch}/d\eta(\eta=0)$ with centrality in the case when $N_{ch}$ 
around mid-pseudo-rapidity is used to determine the centrality, 
as shown in Fig.~\ref{dnchde-w2}b. Also, since the average value 
of $\np$ grows slower with centrality in this case (see Fig.~\ref{b-4way}), 
the difference is further enhanced after dividing $dN_{ch}/d\eta$ by $\np$. 
Figs.~\ref{dnchde-w2} and \ref{npart-nch-w2} thus show that the criteria 
of centrality selection can introduce appreciable differences in the 
centrality dependence of global observables in deuteron-gold collisions, 
and corresponding care must be taken when comparing theoretical 
results with experimental data on centrality dependences.

\begin{acknowledgments}
We thank Ulrich Heinz, Peter Jacobs and Michael Murray for valuable comments. 
This paper is based on work supported by the U.S. Department of Energy under 
Grant No. DE-FG02-01ER41190 (Z.W.L.) and by the U.S. National 
Science Foundation under Grant No. PHY-0098805 as well as the Welch
Foundation under Grant No. A-1358 (C.M.K.)
\end{acknowledgments}


\begin{thebibliography}{99}
\bibitem{hj1382}  
This form of the Hulthen function has been implemented 
in version 1.383 of the HIJING model.
\end{thebibliography}
\end{document}